\documentclass[11pt]{cernrep}
\usepackage{graphicx}
\usepackage{epsfig}
\usepackage{wrapfig}
\usepackage{cite,./mcite}

\newcommand{\pom}{{I\!\!P}}

\newcommand{\xpom}{x_\pom}

\newcommand{\mx}{M_{_{\rm X}}}

\begin{document}

\title{Prospects for {\boldmath $F_L^D$} Measurements at HERA-II}
\author{Paul Newman}
\institute{School of Physics and Astronomy, University of Birmingham,
B15 2TT, United Kingdom}
\maketitle

\begin{abstract}
The theoretical interest in the longitudinal
diffractive structure function $F_L^D$ is briefly motivated and possible
measurement methods are surveyed. A simulation  
based on realistic scenarios with a reduced proton beam energy at
HERA-II using the H1 apparatus shows that measurements are 
possible with up to $4 \sigma$ significance, limited by
systematic errors.
\end{abstract}

\section{Introduction}
\label{sec:intro}

In order to understand
inclusive diffraction fully, it is necessary to separate out 
the contributions from transversely and longitudinally polarised 
exchange photons. 
Here, the formalism of \cite{h1ichep02} is adopted,
where by analogy with inclusive scattering and neglecting weak
interactions, a reduced cross section $\sigma_r^D$ is 
defined,\footnote{It is assumed here that all results are integrated over
$t$. The superscript $(3)$ usually included for $F_2^{D(3)}$ and other 
quantities is dropped for convenience.} related to
the experimentally measured cross section by 
\begin{equation}
\frac{d^3\sigma^{ep \rightarrow eXY}}{{\rm d} x_\pom \ {\rm d} \beta \ {\rm d} Q^2} = \frac{2 \pi \alpha^2}{\beta \, Q^4} \cdot Y_+ \cdot
\sigma_{r}^D(\xpom,\beta,Q^2) \ ,  \ \ \ \ \ {\rm where} \ \ \ \ \  
\sigma_r^D = F_2^D - \frac{y^2}{Y_+} F_L^D 
\label{sigmar}
\end{equation}
and $Y_+ = 1 + (1 - y)^2$. 
The structure function $F_L^D$, is closely related to
the longitudinal photon
contribution, whereas the more familiar $F_2^D$ contains
information on the sum of transverse and longitudinal 
photon contributions. 

It is generally 
understood \cite{Bartels:1998ea} that at high $\beta$ and low-to-moderate 
$Q^2$, $\sigma_r^D$ receives a significant, 
perhaps dominant, higher twist contribution due to longitudinally polarised
photons. Definite predictions \cite{Hebecker:2000xs} 
exist for this contribution, obtained 
by assuming 2-gluon exchange, with a similar phenomenology to
that successfully applied to vector meson cross sections at HERA. 
The dominant role played by gluons in the diffractive
parton densities \cite{h1ichep02} 
implies that the leading twist $F_L^D$ must also be relatively large. 
Assuming the validity of QCD hard scattering collinear
factorisation \cite{Collins:1997sr}, 
this gluon dominance results in a leading twist 
$F_L^D$ which is approximately proportional
to the diffractive gluon density. 
A measurement of $F_L^D$ to even modest
precision would
provide a very powerful
independent tool to verify our understanding of the underlying dynamics
and to test the gluon density extracted indirectly in QCD fits 
from the scaling violations of $F_2^D$.
This is 
particularly
important at the lowest 
$x$ values, where direct information on the gluon density cannot 
be obtained from jet or $D^*$ data due to 
kinematic limitations and where novel effects such as
parton saturation or non-DGLAP dynamics are most likely to become important. 

Several different methods have been proposed to extract information on 
$F_L^D$. It is possible in principle to follow the procedure adopted by H1 in
the inclusive 
case \cite{Adloff:2000qk,h1fl}, exploiting the 
decrease in $\sigma_r^D$
at large $y$ relative to expectations for
$F_2^D$ alone (see equation~\ref{sigmar}). This method may yield significant
results if sufficient precision and $y$ range can be achieved  \cite{jpambw},
though assumptions are required on the $\xpom$ dependence of
$F_2^D$, which is currently not well constrained by
theory. An alternative method, exploiting the azimuthal decorrelation
between the proton and electron scattering planes 
caused by interference between the transverse and longitudinal photon
contributions \cite{Diehl:2005bz}, has already been 
used
with the scattered proton measured in the ZEUS 
LPS \cite{Chekanov:2004hy}. However,
due to the relatively poor statistical precision achievable
with Roman pots at HERA-I, the 
current results are consistent with zero. 
If the potential of the H1 VFPS is fully
realised, this method may yet yield significant results in the HERA-II 
data \cite{vfps}.
However, if the necessary data are taken, the most promising possibility
is to extract $F_L^D$ by comparing
data at the same $Q^2$, $\beta$ and $\xpom$, but from different centre 
of mass energies $\surd s$ and hence from different $y$ values. 
The longitudinal
structure function can then be extracted directly and model-independently 
from the measured data using equation~\ref{sigmar}.
In this contribution, one possible scenario 
is investigated, based
on modified beam energies and luminosities which are currently under discussion
as a possible part of the HERA-II programme.

\section{Simulated {\boldmath $F_L^D$} Measurement}

Given the need to obtain a large integrated luminosity at the
highest possible beam energy for the remainder of the HERA programme and
the fixed end-point in mid 2007, it is likely 
that only a relatively
small amount of data can be taken with reduced beam energies.
A possible scenario is investigated here in which $10 \ {\rm pb^{-1}}$ are
taken
at just one
reduced proton beam energy of $E_p = 400 \ {\rm GeV}$, the electron beam
energy being unchanged at $27.5 \ {\rm GeV}$. 
Since the maximum achievable 
instantaneous luminosity at HERA scales like the proton beam energy
squared \cite{willeke}, this data sample
could be obtained in around 2-3 months at the current level of HERA
performance.
It is assumed that a larger data
volume of $100 \ {\rm pb^{-1}}$ is available at $E_p = 920 \ {\rm GeV}$,
which allows for downscaling of high rate
low $Q^2$ inclusive triggers.\footnote{Alternative scenarios in which
a smaller data volume at large $E_p$ is taken in a short, dedicated
run, could potentially lead to better controlled systematics at the expense
of increased statistical errors.} 
The results presented here can be used to infer those from other 
scenarios given that the statistical uncertainty scales like
$\sigma_r^{D \, 400} / \sqrt{{\cal L}_{400}} +
\sigma_r^{D \, 920} / \sqrt{{\cal L}_{920}}$, 
where $\sigma_r^{D \, E_p}$ and
${\cal L}_{E_p}$ are the reduced cross section and the luminosity at a 
proton beam energy of $E_p$, respectively.

\begin{figure}[h] \unitlength 1mm
 \begin{center}
 \begin{picture}(100,65)

  \put(-25,-10){\epsfig{file=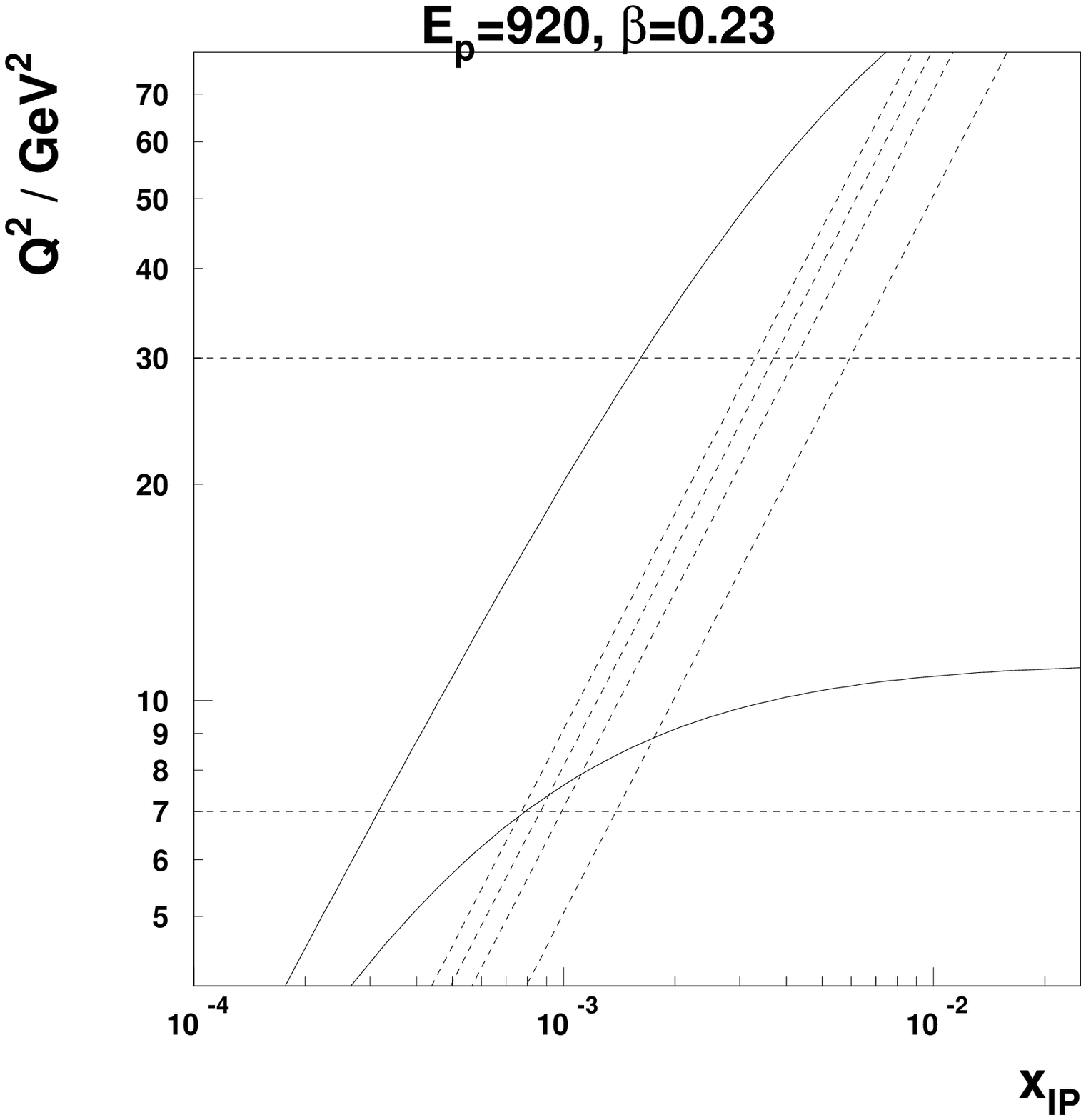,width=0.47\textwidth}}
  \put(60,-10){\epsfig{file=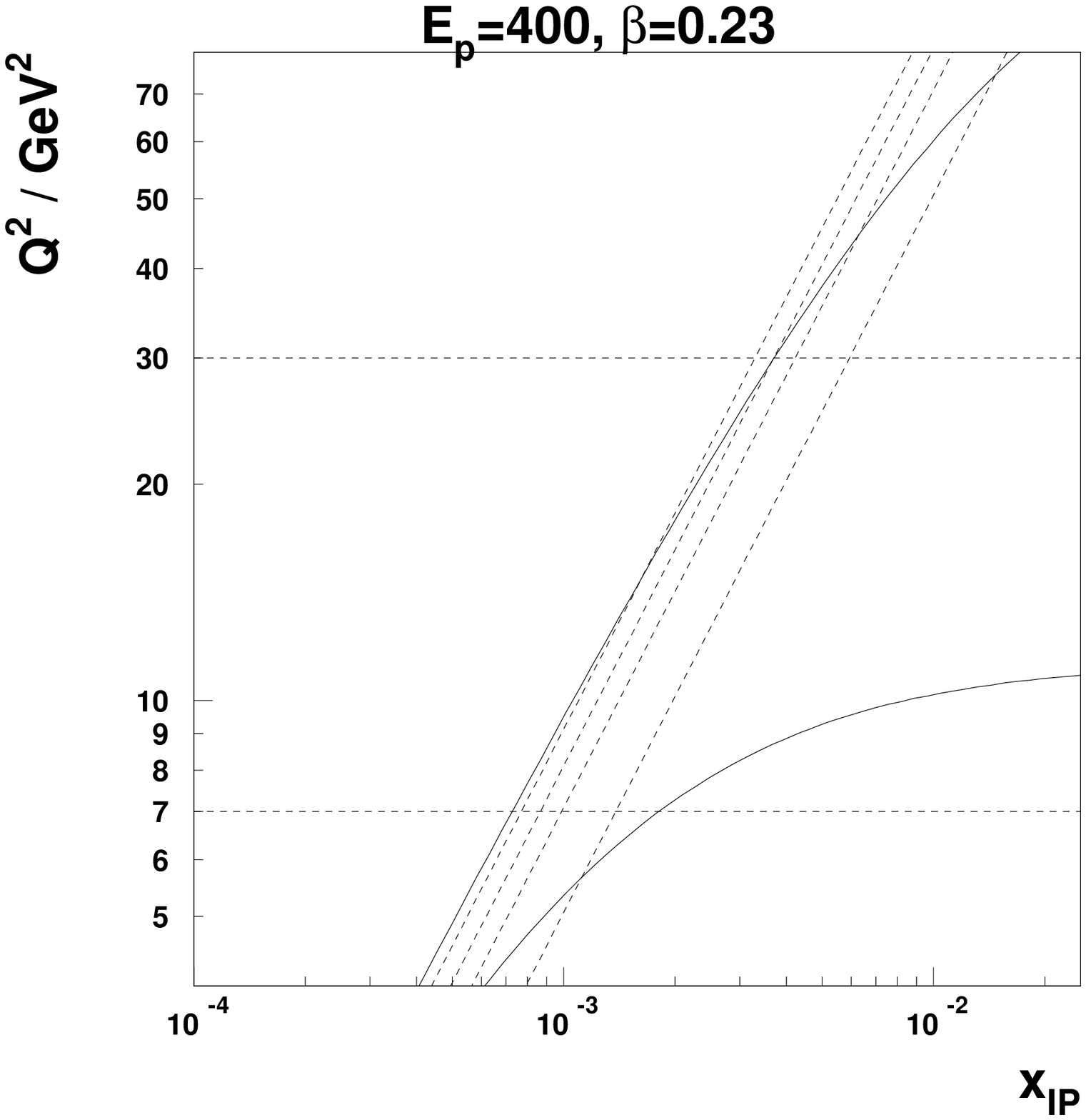,width=0.47\textwidth}}
 \end{picture}
 \end{center}
 \caption{Illustration of the kinematic plane in $Q^2$ and $\xpom$
at proton energies of $920 \ {\rm GeV}$ and $400 \ {\rm GeV}$, with
fixed $\beta = x / \xpom = 0.23$. The solid lines illustrate the experimental
limits of 
$155^\circ < \theta_e^\prime < 173^\circ$. The horizontal dashed lines
illustrate the $Q^2$ range used for the simulation. The diagonal dashed
lines illustrate the binning in $y$, corresponding at $E_p = 400 \ {\rm GeV}$
to $y = 0.9$ (leftmost line), $y = 0.8$, $y=0.7$ and $y = 0.5$ 
(rightmost line).}
\label{kinplane}
\end{figure}

The longitudinal structure function can be extracted 
from the data at the two beam energies using
\begin{equation}
  F_L^D = 
\frac{Y_+^{400} \ Y_+^{920}}{y^2_{400} Y_+^{920} - y^2_{920} Y_+^{400}}
\ \left( \sigma_r^{D \, 920} - \sigma_r^{D \, 400} \right) \ ,
\label{eqn:fld}
\end{equation}
where $y_{E_p}$ and $Y_+^{E_p}$ denote $y$ and $Y_+$ at a beam energy $E_p$. 
It is clear from equation~\ref{eqn:fld} that the best sensitivity to $F_L^D$
requires the maximum difference between the reduced cross sections at the
two beam energies, which (equation~\ref{sigmar}) implies the maximum
possible $y$ at $E_p = 400 \ {\rm GeV}$. By measuring scattered electrons 
with energies $E_e^\prime$
as low as $3 \ {\rm GeV}$ \cite{Adloff:2000qk}, 
the H1 collaboration has obtained data at
$y = 0.9$. This is possible with the 
use of the SPACAL calorimeter in
combination with a measurement of the electron track in either 
the backward silicon tracker (BST) or the central jet chamber (CJC).
For HERA-II running,
the corresponding available range of scattered electron polar angle
is $155^\circ < \theta_e^\prime < 173^\circ$,
which is used in the current 
study.\footnote{One interesting alternative
running scenario \cite{Klein:2004zq} is to
obtain data at $E_p = 920 \ {\rm GeV}$ with the vertex shifted by 
$20 \ {\rm cm}$ in the outgoing proton direction, which would allow
measurements up to $\theta_e^\prime = 175^\circ$, giving a low $Q^2$ acceptance
range which closely matches that for the $E_p = 400 \ {\rm GeV}$ data
at the normal vertex position.}
Three intervals in $y$ are considered, corresponding at 
$E_p = 400 \ {\rm GeV}$ to $0.5 < y_{400} < 0.7$, $0.7 < y_{400} < 0.8$ and
$0.8 < y_{400} < 0.9$. It is ensured that identical ranges in $\beta$,
$\xpom$ and $Q^2$ 
are studied at $E_p = 920 \ {\rm GeV}$ by choosing the bin edges such that
$y_{920} = y_{400} \cdot 400 / 920$.
Since the highest possible precision is required in
this measurement, the restriction $\xpom < 0.02$ is imposed, which leads to
negligible acceptance losses with a typical cut on the forwardmost 
extent of the diffractive system $\eta_{\rm max} < 3.3$. The kinematic
restrictions on $E_e^\prime$, 
$\theta_e^\prime$ and $\xpom$
lead to almost no change in the mean $Q^2$, $\mx^2$ or
$\beta \simeq Q^2 / (Q^2 + \mx^2)$ as either 
$y$ or $E_p$ are varied. In contrast,
$\xpom = Q^2 / (s \, y \, \beta)$ varies approximately as $1 / y$. 
As is shown in figure~\ref{kinplane}, at the average $\beta = 0.23$,
there is at least partial acceptance for all $y$ bins in the range
$7 < Q^2 < 30 \ {\rm GeV^2}$, which is chosen for this study, leading to
an average value of $Q^2$ close to $12 \ {\rm GeV^2}$.

The simulation is performed using the 
RAPGAP \cite{Jung:1993gf} Monte Carlo generator
to extract the number of events per unit luminosity in each bin at each
centre of mass energy. 
The values of $F_2^D$ and $F_L^D$, and hence $\sigma_r^{D \, 920}$
and $\sigma_r^{D \, 400}$ are obtained using an updated version
of the preliminary H1 2002 NLO QCD fit \cite{h1ichep02}. 

The expected precision on $F_L^D$ is 
obtained by error propagation through equation~\ref{eqn:fld}.
The systematic uncertainties are estimated on the
basis of previous experience with the H1 
detector \cite{h1ichep02,Adloff:2000qk}. 
At the large $y$ values involved, the 
kinematic variables are most accurately reconstructed using the electron
energy and angle alone. The systematic 
uncertainties on the measurements of these quantities are assumed
to be correlated between the two beam energies. 
With the use of the BST and CJC, the possible bias in the measurement of 
$\theta_e^\prime$ is at the level of $0.2 \ {\rm mrad}$. The
energy scale of the SPACAL calorimeter is known with a precision varying
linearly from 2\% at $E_e^\prime = 3 \ {\rm GeV}$ to 0.2\% at 
$E_e^\prime = 27.5 \ {\rm GeV}$.
Other uncertainties which are correlated between the two beam energies
arise from the photoproduction background subtraction (important
at large $y$ and assumed to be known with a precision of $25\%$) and the 
energy scale for the hadronic final state used in the reconstruction
of $\mx$ and hence $\xpom$ (taken to
be known to 4\%, as currently). Sources of uncertainty 
which are assumed 
to be uncorrelated between the low and high $E_p$ measurements
are
the luminosity measurement (taken to be $\pm 1 \%$), the trigger and 
electron track
efficiencies ($\pm 1 \%$ combined) and the acceptance corrections, 
obtained using
RAPGAP ($\pm 2 \%$). The combined uncorrelated error is thus $2.4 \%$.
Finally, a normalisation uncertainty of $\pm 6 \%$
due to corrections for proton dissociation contributions is taken to
act simultaneously in the two measurements. 
Other sources of uncertainty currently considered in
H1 measurements of diffraction are negligible in the kinematic
region studied here. 

\begin{table}[h]
\begin{center}
{\footnotesize
\begin{tabular}{|l|l|l||l|l||l|l|l|l|l|l||l|l|l|}
\hline
$y_{400}$ & $y_{920}$ & $\xpom$ & $F_2^D$ & $F_L^D$ &
$\delta_{\rm unc}$ & $\delta_{\rm norm}$ &
$\delta E_e^\prime$ & $\delta \theta_e^\prime$ & $\delta \mx$ & 
$\delta \gamma p$ & $\delta_{\rm syst}$ & 
$\delta_{\rm stat}$ & $\delta_{\rm tot}$  \\ \hline
$0.5 - 0.7$ & 0.217 - 0.304 & 0.0020 & 15.72 & 3.94 & 34 & 6 & 8 & 2 & 7 & 0  & 36 & 20 & 41  \\ \hline
$0.7 - 0.8$ & 0.304 - 0.348 & 0.0016 & 20.87 & 5.25 & 19 & 6 & 3 & 2 & 5 & 6  & 22 & 17 & 28  \\ \hline
$0.8 - 0.9$ & 0.348 - 0.391 & 0.0014 & 24.47 & 6.16 & 14 & 6 & 6 & 1 & 2 & 13 & 21 & 13 & 25  \\ \hline
\end{tabular}
}
\end{center}
\caption{Summary of the simulation at
$Q^2 = 12 \ {\rm GeV}$ and $\beta = 0.23$. The first three columns
contain the $y$ ranges used at 
$E_p = 400 \ {\rm GeV}$ and $E_p = 920 \ {\rm GeV}$ and the $\xpom$ values.
The next two columns contain the values of the diffractive
structure functions. These are followed by the uncorrelated
($\delta_{\rm unc}$) and
proton dissociation ($\delta_{\rm norm}$) uncertainties
and the correlated systematics due to the
electron energy ($\delta E_e^\prime$)
and angle ($\delta \theta_e^\prime$)
measurements, the hadronic energy scale ($\delta \mx$) and the
photoproduction background ($\delta \gamma p$), all in percent. 
The last three columns summarise
the systematic, statistical and total uncertainties.}
\label{table:fld}
\end{table}

\begin{wrapfigure}{L}{0.41\linewidth}
\centering
\epsfig{file=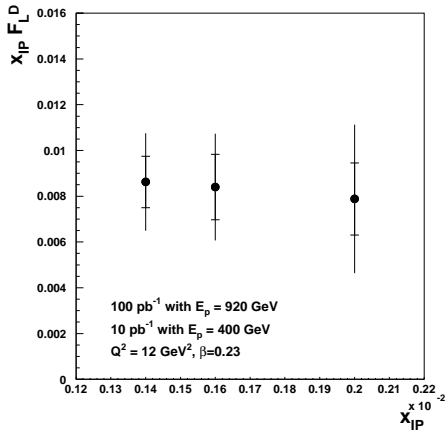,width=0.99\linewidth}
\caption{Illustration of the 
simulated result for $F_L^D$, showing the three data
points with statistical (inner bars) and total (outer bars) errors.}
\label{fig:fldshow}
\end{wrapfigure}

Full details of the 
simulated uncertainties on the $F_L^D$ measurements are given in 
table~\ref{table:fld}. An illustration of the corresponding expected
measurement, based on the $F_L^D$ from the H1 2002 fit is shown in 
figure~\ref{fig:fldshow}. The most precise
measurement is obtained at the highest $y$, 
where $F_L^D$ would be 
determined to be unambiguously different from its maximum value of $F_2^D$ 
and to be non-zero at the $4 \sigma$ level.
Two further measurements
are obtained at lower $y$ values. The dominant 
errors arise from statistical uncertainties and from
uncertainties which are uncorrelated between the two
beam energies. Minimising the latter is a major experimental
challenge to be addressed in the coming years. 

Only one possible scenario has been investigated here, leading to a 
highly encouraging result at relatively low $\beta$, which would provide
a very good test of the leading twist $F_L^D$ and thus of the gluon
density extracted in QCD fits to $F_2^D$. 
It may also be possible to obtain results at high $\beta$, giving 
information on the higher twist contributions in that region,
for example by restricting the analysis to lower $\xpom$.

\vspace*{-0.5cm}
\section*{Acknowledgements}

For comments, corrections, cross-checks
and code, thanks to Markus Diehl, Joel Feltesse, Max Klein
and Frank-Peter Schilling!

\bibliographystyle{heralhc} 
{\raggedright
\bibliography{fld}
}
\end{document}